\documentclass[aps,twocolumn,twoside,nofootinbib,preprintnumbers]{revtex4}

\usepackage{graphicx}
\usepackage{amsmath}
\newcommand{\<}{\langle}
\renewcommand{\>}{\rangle}

\newcommand{\be}{\begin{equation}}
\newcommand{\ee}{\end{equation}}
\def\ba#1\ea{\begin{align}#1\end{align}} % workaround for amsmath
\newcommand{\nn}{\nonumber\\}

\newcommand{\cond}[1]{\left\{\begin{array}{l@{~~~}l}#1\end{array}\right.}

\newcommand{\E}{{\cal E}}
\renewcommand{\d}{{\mathrm{d}}}

\newcommand{\poly}{\mathop{\mathrm{poly}}}

\begin{document}

\title{Spatial search by quantum walk}

\author{Andrew M. Childs}
\email[]{amchilds@mit.edu}

\author{Jeffrey Goldstone}
\email[]{goldston@mit.edu}

\affiliation{Center for Theoretical Physics \\
             Massachusetts Institute of Technology \\
             Cambridge, MA 02139, USA}

\preprint{MIT-CTP \#3384}

%%%%%%%%%%%%%%%%%%%%%%%%%%%%%%%%%%%%%%%%%%%%%%%%%%%%%%%%%%%%%%%%%%%%%%%%%%%%%%%
% Abstract

\begin{abstract}
Grover's quantum search algorithm provides a way to speed up
combinatorial search, but is not directly applicable to searching a
physical database.  Nevertheless, Aaronson and Ambainis showed that a
database of $N$ items laid out in $d$ spatial dimensions can be searched
in time of order $\sqrt N$ for $d>2$, and in time of order $\sqrt N
\poly(\log N)$ for $d=2$.  We consider an alternative search algorithm
based on a continuous time quantum walk on a graph.  The case of the
complete graph gives the continuous time search algorithm of Farhi and
Gutmann, and other previously known results can be used to show that
$\sqrt N$ speedup can also be achieved on the hypercube.  We show that
full $\sqrt N$ speedup can be achieved on a $d$-dimensional periodic
lattice for $d>4$.  In $d=4$, the quantum walk search algorithm takes
time of order $\sqrt N \poly(\log N)$, and in $d<4$, the algorithm
does not provide substantial speedup.
\end{abstract}

\maketitle

%%%%%%%%%%%%%%%%%%%%%%%%%%%%%%%%%%%%%%%%%%%%%%%%%%%%%%%%%%%%%%%%%%%%%%%%%%%%%%%
\section{Introduction} 
\label{sec:intro}

Grover's quantum search algorithm \cite{Gro97} is one of the main
applications of quantum computation.  Given a black box function
$f(x):\{1,\ldots,N\} \to \{0,1\}$ satisfying
\be
  f(x) = \cond{0 & x \ne w \\ 
               1 & x=w \,,}
\ee
Grover's algorithm can find the value of $w$ using of order $\sqrt N$
queries, which is optimal \cite{BBBV97}.  On the other hand, no
classical algorithm can do better than exhaustive search, which takes of
order $N$ queries.  Therefore Grover's algorithm can be used to speed up
brute force combinatorial search.  It can also be used as a subroutine
in a variety of other quantum algorithms.

Grover's algorithm is sometimes described as a way to search an unsorted
database of $N$ items in time $O(\sqrt N)$.  But the algorithm as
originally proposed is not designed to search a physical database.
Suppose we had $N$ items stored in a $d$-dimensional physical space, and
that these items could be explored in superposition by a quantum
computer making local moves (a ``quantum robot'' \cite{Ben02}).
Naively, it would seem that each step of the Grover algorithm should
take time of order $N^{1/d}$, since this is the time required to cross
the database.  Performing $\sqrt N$ iterations, we find that the search
takes time of order $N^{\frac{1}{2}+\frac{1}{d}}$, so no speedup is
achieved in $d=2$, and full speedup is achieved only in the limit of
large $d$.

However, it is possible to do better than this naive approach suggests.
In \cite{AA03}, Aaronson and Ambainis present a model of query
complexity on graphs.  Within this model, they give a recursive
algorithm for the search problem that achieves full $\sqrt N$ speedup
for a $d \ge 3$ dimensional lattice, and runs in time $\sqrt N \,
\log^2 N$ in $d=2$.  (It is obvious that no algorithm can get speedup
in $d=1$.)

In this paper we approach the spatial search problem using quantum
walks.  Since random walks are commonly used in classical algorithms, it
is natural to consider a quantum analogue of a classical random walk as
an algorithmic tool.  Here we consider the continuous time quantum walk
\cite{FG98}.  On certain graphs, this quantum walk can yield
exponentially faster hitting times than its classical counterpart
\cite{FG98,CFG02}.  Indeed, a recent result shows that the continuous
time quantum walk can solve a certain black box problem exponentially
faster than any classical algorithm \cite{CCDFGS03}.

Quantum walks provide a natural framework for the spatial search
problem because the graph can be used to model the locality of the
database.  We present a simple quantum walk search algorithm that can
be applied to any graph.  Our algorithm could be implemented within
the model of \cite{AA03}, but is actually much simpler because it uses
no auxiliary storage space.  For the case of the complete graph, the
resulting algorithm is simply the continuous time search algorithm of
Farhi and Gutmann \cite{FG96}.  On the hypercube, previous results can
be used to show that the algorithm also provides quadratic speedup
\cite{FGGS00,CDFGGL02}.  However, in both of these cases, the graph is
highly connected.  Here, we consider the case of a $d$-dimensional
cubic periodic lattice, where $d$ is fixed independent of $N$.  We
find full $\sqrt N$ speedup in $d > 4$ and running time $O(\sqrt N
\log^{3/2} N)$ in $d=4$.  In $d < 4$, we find that quadratic speedup
is impossible, so the continuous time quantum walk algorithm is never
faster than the Aaronson-Ambainis algorithm.

We note that it is also possible to construct a quantum analogue of a
discrete time random walk \cite{AAKV01,NV01} (although the walk cannot
take place directly on the vertices of the graph \cite{Mey96}).  This
type of walk has been used to construct a fast search algorithm on the
hypercube \cite{SKW02}, and more recently, on a $d$-dimensional
lattice with $d \ge 2$ \cite{AKR}.  The latter result outperforms our
continuous-time walk algorithm for $d=2,3,4$.  However, similar
performance can be achieved by a modification of the continuous-time
algorithm \cite{CG04}.

This paper is organized as follows.  In Section \ref{sec:qwalk} we
review the continuous time quantum walk and show how it can be used to
approach the search problem.  In Section \ref{sec:highdim} we review
the results in the high dimensional cases (the complete graph and the
hypercube), casting them in the language of continuous time quantum
walks.  In Section \ref{sec:finite} we present the results for finite
dimensional lattices, and in Section \ref{sec:discussion}, we conclude
with a discussion of our results.

%%%%%%%%%%%%%%%%%%%%%%%%%%%%%%%%%%%%%%%%%%%%%%%%%%%%%%%%%%%%%%%%%%%%%%%%%%%%%%%
\section{Quantum walk} 
\label{sec:qwalk}

The continuous time quantum walk on a graph is defined in direct
analogy to a continuous time classical random walk \cite{FG98}.  Given
an undirected graph $G$ with $N$ vertices and no self loops, we define
the {\em adjacency matrix}
\be
  A_{jk} = \cond{1 & (j,k) \in G \\
                 0 & {\rm otherwise}}
\ee
which describes the connectivity of $G$.  In terms of this matrix, we
can also define the {\em Laplacian} $L = A - D$, where $D$ is the
diagonal matrix with $D_{jj} = \deg(j)$, the degree of vertex $j$.
The continuous time random walk on $G$ is a Markov process with a
fixed probability per unit time $\gamma$ of jumping to an adjacent
vertex.  In other words, the probability of jumping to any connected
vertex in a time $\epsilon$ is $\gamma \epsilon$ (in the limit
$\epsilon \to 0$).  This walk can be described by the first-order,
linear differential equation
\be
  \frac{\d p_j(t)}{\d t} = \gamma \sum_k L_{jk} \, p_k(t)
\,,
\label{eq:markov}
\ee
where $p_j(t)$ is the probability of being at vertex $j$ at time $t$.
Since the columns of $L$ sum to zero, probability is conserved.

The continuous time quantum walk on a graph takes place in an
$N$-dimensional Hilbert space spanned by states $|j\>$, where $j$ is a
vertex in $G$.  In terms of these basis states, we can write a general
state $|\psi(t)\>$ in terms of the $N$ complex amplitudes $q_j(t) =
\<j|\psi(t)\>$.  If the Hamiltonian is $H$, then the dynamics of the
system are determined by the Schr\"odinger equation,\footnote{We have
chosen units in which $\hbar=1$.}
\be
  i \frac{\d q_j(t)}{\d t} = \sum_k H_{jk} \, q_k(t)
\,.
\label{eq:schrodinger}
\ee Note the similarity between (\ref{eq:markov}) and
(\ref{eq:schrodinger}).  The continuous time quantum walk is defined
by simply letting $H=-\gamma L$.\footnote{Here the sign is chosen so
that the Hamiltonian is positive semidefinite.  We have defined
$L=A-D$ so that for a lattice, $L$ is a discrete approximation to the
continuum operator $\nabla^2$.  A free particle in the continuum has
the positive semidefinite Hamiltonian $H=-\nabla^2$ (in appropriate
units).}  Then the only difference between (\ref{eq:markov}) and
(\ref{eq:schrodinger}) is a factor of $i$, which nevertheless can
result in radically different behavior.

As an aside, we note that the Laplacian does not provide the only
possible Hamiltonian for a quantum walk.  Whereas (\ref{eq:markov})
requires $\sum_j L_{jk} = 0$ to be a valid probability-conserving
classical Markov process, (\ref{eq:schrodinger}) requires $H=H^\dag$
to be a valid unitary quantum process.  Therefore we could also
choose, for example, $H=-\gamma A$.  All of the graphs we consider in
this paper are regular (i.e., $\deg(j)$ is independent of $j$), so
these two choices give rise to the same quantum dynamics.  However,
for non-regular graphs the two choices will give different results.

To approach the Grover problem with a quantum walk, we need to modify
the Hamiltonian so that the vertex $w$ is special.  Following
\cite{FG96}, we introduce the {\em oracle Hamiltonian}\footnote{More
precisely, we should use $H_w = -\omega |w\>\<w|$ where $\omega$ is a
fixed parameter with units of inverse time.  However, we choose units
in which $\omega=1$.  In these units, $\gamma$ is a dimensionless
parameter.}
\be
  H_w = -|w\>\<w|
\ee
which has energy zero for all states except $|w\>$, which is the
ground state, with energy $-1$.  Solving the Grover problem is
equivalent to finding the ground state of this Hamiltonian.  In this
paper we assume that this Hamiltonian is given, and we want to use it
for as little time as possible to find the value of $w$.  Note that
this Hamiltonian could be simulated in the circuit model using the
standard Grover oracle
\be
  U_w |j\> = (-1)^{\delta_{jw}} |j\>
\,.
\label{eq:oracle}
\ee
However, in this paper we focus on the continuous time description.

To construct an algorithm with the locality of a particular graph $G$,
we consider the time-independent Hamiltonian
\be
  H = -\gamma L + H_w = -\gamma L - |w\>\<w|
\ee
where $L$ is the Laplacian of $G$.  We begin in a uniform
superposition over all vertices of the graph,
\be
  |s\> = \frac{1}{\sqrt N} \sum_j |j\>
\,,
\ee
and run the quantum walk for time $T$.  We then measure in the vertex
basis.  Our objective is to choose the parameter $\gamma$ so that the
success probability $|\<w|\psi(T)\>|^2$ is as close to $1$ as possible
for as small a $T$ as possible.  Note that the coefficient of $H_w$ is
held fixed at $1$ to make the problem fair (e.g., so that evolution for
time $T$ could be simulated with $O(T)$ queries of the standard Grover
oracle (\ref{eq:oracle})).

One might ask why we should expect this algorithm to give a
substantial success probability for some values of $\gamma,T$.  We
motivate this possibility in terms of the spectrum of $H$.  Note that
regardless of the graph, $|s\>$ is the ground state of the Laplacian,
with $L|s\>=0$.  As $\gamma \to \infty$, the contribution of $H_w$ to
$H$ is negligible, so the ground state of $H$ is close to $|s\>$.  On
the other hand, as $\gamma \to 0$, the contribution of $L$ to $H$
disappears, so the ground state of $H$ is close to $|w\>$.
Furthermore, since $|s\>$ is nearly orthogonal to $|w\>$, degenerate
perturbation theory shows that the first excited state of $H$ will be
close to $|s\>$ as $\gamma \to 0$ for large $N$.  We might expect that
over some intermediate range of $\gamma$, the ground state will switch
from $|w\>$ to $|s\>$, and could have substantial overlap on both for
a certain range of $\gamma$.  If the first excited state also has
substantial overlap on both $|w\>$ and $|s\>$ at such values of
$\gamma$, then the Hamiltonian will drive transitions between the two
states, and thus will rotate the state from $|s\>$ to a state with
substantial overlap with $|w\>$ in a time of order $1/(E_1-E_0)$,
where $E_0$ is the ground state energy and $E_1$ is the first excited
state energy.

Indeed, we will see that this is a good description of the algorithm
if the dimension of the graph is sufficiently high.  The simplest example
is the complete graph (the ``analog analogue'' of the Grover algorithm
\cite{FG96}) which can be thought of roughly as having dimension
proportional to $N$.  A similar picture holds for the $(\log
N)$-dimensional hypercube.  When we consider a $d$-dimensional lattice
with $d$ independent of $N$, we will see that the state $|s\>$ still
switches from ground state to first excited state at some critical
value of $\gamma$.  However, the $|w\>$ state does not have
substantial overlap on the ground and first excited states unless
$d>4$, so the algorithm will not work for $d<4$ (and $d=4$ will be a
marginal case).

%%%%%%%%%%%%%%%%%%%%%%%%%%%%%%%%%%%%%%%%%%%%%%%%%%%%%%%%%%%%%%%%%%%%%%%%%%%%%%%
\section{High dimensions} 
\label{sec:highdim}

In this section, we describe the quantum walk algorithm on ``high
dimensional'' graphs, namely the complete graph and the hypercube.
These cases have been analyzed in previous works
\cite{FG96,FGGS00,CDFGGL02}.  Here, we reinterpret them as quantum
walk algorithms, which provides motivation for the case of a lattice
in $d$ spatial dimensions.

%%%%%%%%%%%%%%%%%%%%%%%%%%%%%%%%%%%%%%%%%%%%%%%%%%%%%%%%%%%%%%%%%%%%%%%%%%%%%%%
\subsection{Complete graph}

Letting $L$ be the Laplacian of the complete graph, we find exactly the
continuous time search algorithm proposed in \cite{FG96}.  Adding a
multiple of the identity matrix to the Laplacian gives
\be
   L + N I = N|s\>\<s| 
           = \begin{pmatrix}1      & \cdots & 1      \cr
                            \vdots & \ddots & \vdots \cr
                            1      & \cdots & 1\end{pmatrix}
\,.
\ee
Therefore we consider the Hamiltonian
\be
  H = -\gamma N |s\>\<s| - |w\>\<w|
\,.
\ee

\begin{figure}
\includegraphics[width=\columnwidth]{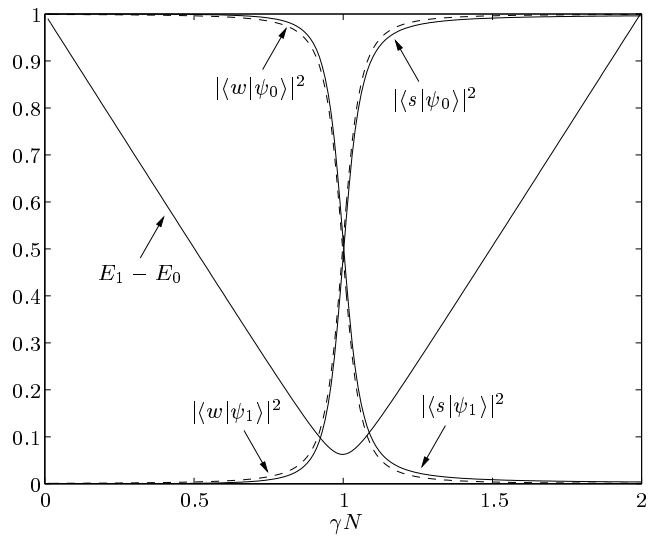}
\caption{Energy gap and overlaps for the complete graph with $N=1024$.}
\label{fig:completegraph}
\end{figure}

Since this Hamiltonian acts nontrivially only on a two-dimensional
subspace, it is straightforward to compute its spectrum exactly for
any value of $\gamma$.  For $\gamma N \ll 1$, the ground state is
close to $|w\>$, and for $\gamma N \gg 1$, the ground state is close to
$|s\>$.  In fact, for large $N$, there is a sharp change in the ground
state from $|w\>$ to $|s\>$ as $\gamma N$ is varied from slightly less
than $1$ to slightly greater than $1$.  Correspondingly, the gap between
the ground and first excited state energies is smallest for $\gamma N
\sim 1$, as shown in Figure \ref{fig:completegraph}.  At $\gamma N=1$,
for $N$ large, the eigenstates are $\frac{1}{\sqrt 2} (|w\> \pm |s\>)$
(up to terms of order $N^{-1/2}$), with a gap of $2 / \sqrt N$.  Thus
the walk rotates the state from $|s\>$ to $|w\>$ in time $\pi \sqrt N
/ 2$.

%%%%%%%%%%%%%%%%%%%%%%%%%%%%%%%%%%%%%%%%%%%%%%%%%%%%%%%%%%%%%%%%%%%%%%%%%%%%%%%
\subsection{Hypercube}

Now consider the $n$-dimensional hypercube with $N=2^n$ vertices.  The
vertices of the graph are labeled by $n$-bit strings, and two
vertices are connected if and only if they differ in a single bit.
Therefore the adjacency matrix can be written as
\be
  A = \sum_{j=1}^n \sigma_x^{(j)}
\ee
where $\sigma_x^{(j)}$ is the Pauli sigma $x$ operator on the $j$th
bit.

\begin{figure}
\includegraphics[width=\columnwidth]{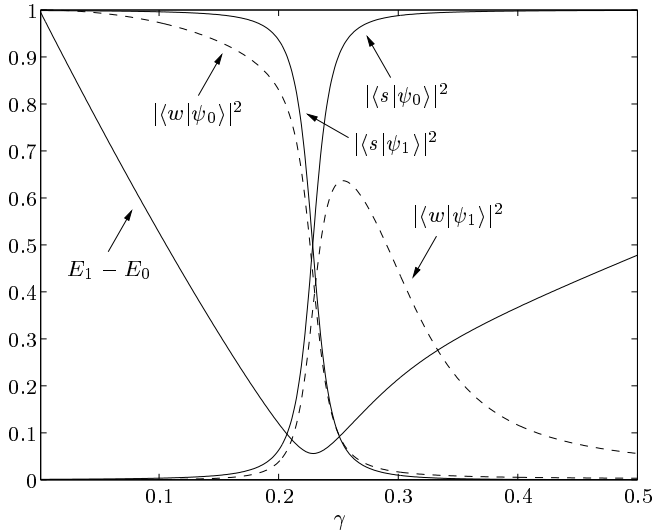}
\caption{Energy gap and overlaps for the hypercube with $N=2^{10}=1024$.}
\label{fig:hypercube}
\end{figure}

In this case, we again find a sharp transition in the eigenstates at a
certain critical value of $\gamma$, as shown in Figure
\ref{fig:hypercube}.  The Hamiltonian can be analyzed using
essentially the same method we will apply in the next section,
together with facts about spin operators.  The energy gap is analyzed
in Section 4.2 of \cite{FGGS00}, and the energy eigenstates are
analyzed in Appendix B of \cite{CDFGGL02}.  The critical value of
$\gamma$ is
\be
  \gamma=\frac{1}{2^n} \sum_{r=1}^n \binom{n}{r} \frac{1}{r}
        =\frac{2}{n} + O(n^{-2})
\,,
\ee
at which the energy gap is
\be
  {2 \over \sqrt N} [1 + O(n^{-1})]
\ee
and the ground and first excited states are ${1 \over \sqrt2}(|w\> \pm
|s\>)$ up to terms of order $1/n$.  Again, we find that after a time
of order $\sqrt N$, the probability of finding $w$ is of order $1$.

%%%%%%%%%%%%%%%%%%%%%%%%%%%%%%%%%%%%%%%%%%%%%%%%%%%%%%%%%%%%%%%%%%%%%%%%%%%%%%%
\section{Finite dimensions} 
\label{sec:finite}

\begin{figure}
\includegraphics[width=.748\columnwidth]{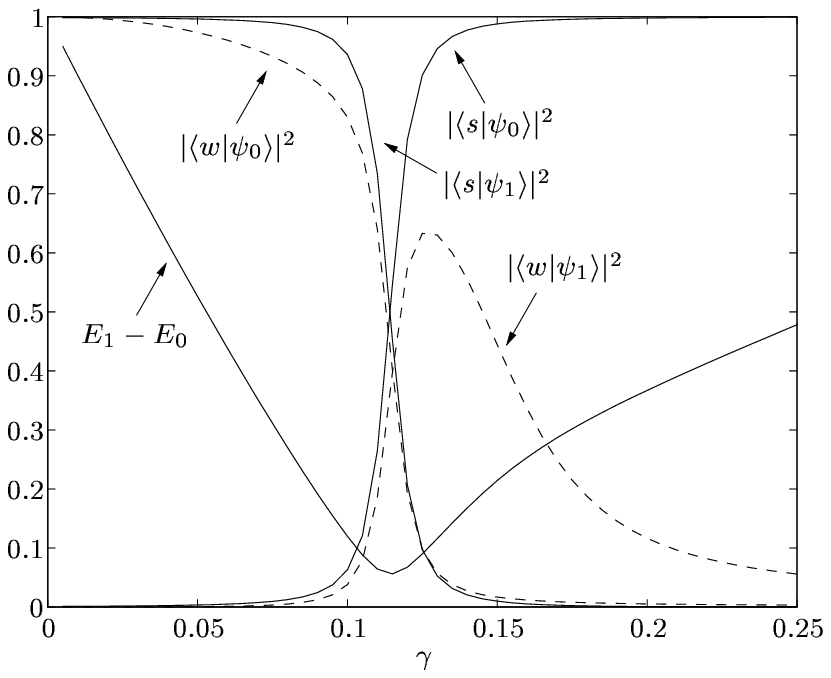} \\
\includegraphics[width=.748\columnwidth]{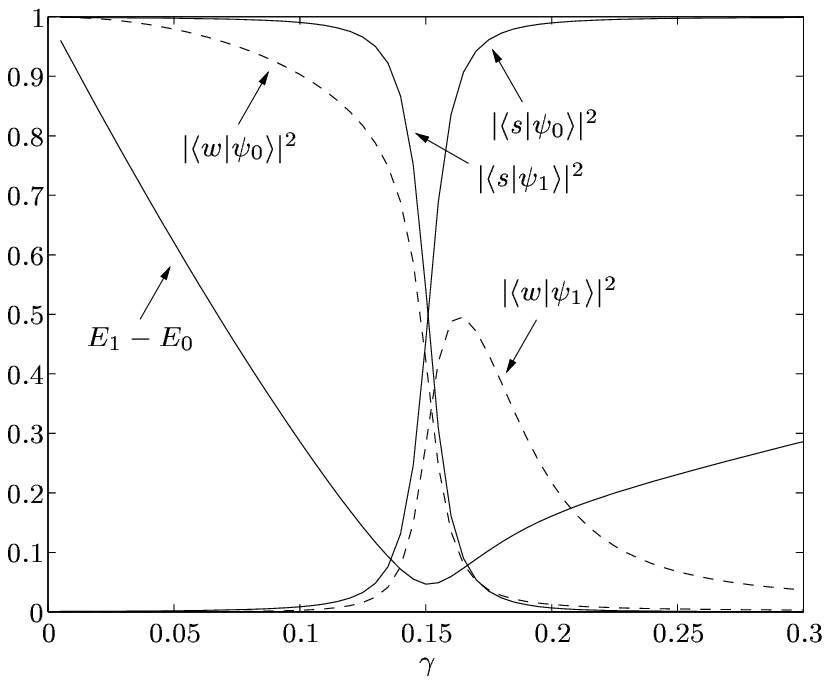} \\
\includegraphics[width=.748\columnwidth]{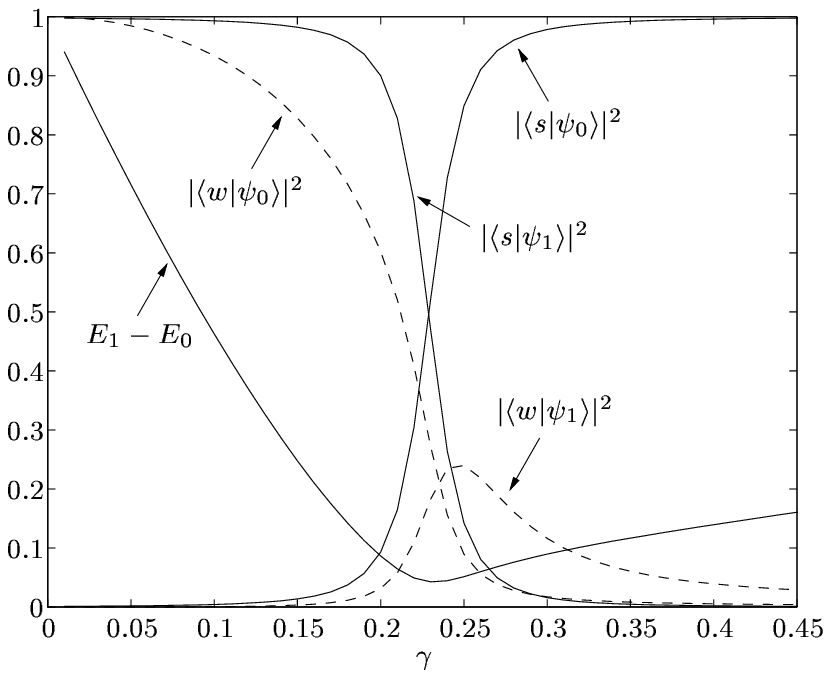} \\
\includegraphics[width=.748\columnwidth]{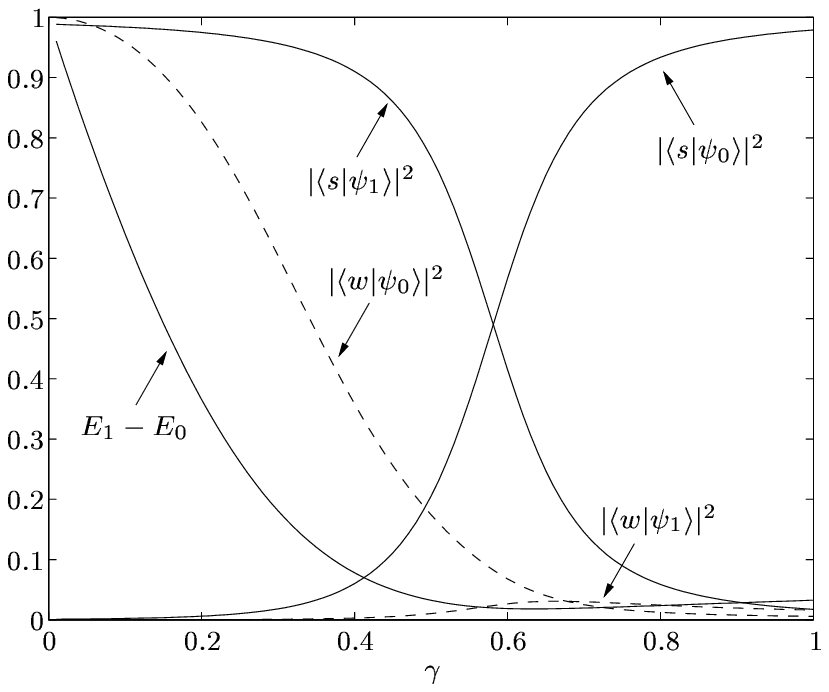}
\caption{Energy gap and overlaps for $d$-dimensional lattices with $N
\approx 1000$.  From top to bottom, $d=5$, $N=4^5=1024$; $d=4$,
$N=6^4=1296$; $d=3$, $N=10^3=1000$; $d=2$, $N=32^2=1024$.}
\label{fig:ddim}
\end{figure}

Having seen that the algorithm works in two cases where the dimension
of the graph grows with $N$, we now consider the case of a $d$
dimensional cubic periodic lattice, where $d$ is fixed independent of
$N$.  The minimum gap and overlaps of $|s\>,|w\>$ with the ground and
first excited states are shown in Figure \ref{fig:ddim} for
$d=2,3,4,5$ and $N \approx 1000$.  In all of these plots, there is a
critical value of $\gamma$ where the energy gap is a minimum, and in
the vicinity of this value, the state $|s\>$ changes from being the
first excited state to being the ground state.  In large enough $d$,
the $|w\>$ state changes from being the ground state to having large
overlap on the first excited state in the same region of $\gamma$.
However, for smaller $d$, the range of $\gamma$ over which the change
occurs is wider, and the overlap of the $|w\>$ state on the lowest two
eigenstates is smaller.  Note that in all cases, $|s\>$ is supported
almost entirely on the subspace of the two lowest energy states.
Therefore, if the algorithm starting in the state $|s\>$ is to work at
all, it must work essentially in a two dimensional subspace.

In the rest of this section, we will make this picture quantitative.
We begin with some general techniques for analyzing the spectrum of
$H$ using knowledge of the spectrum of the graph.  We then show the
existence of a phase transition in $\gamma$, and we show that for any
$d$, the algorithm fails if $\gamma$ is not close to a certain
critical value.
Next we consider what happens when $\gamma$ is close to its critical
value.  In $d > 4$, we show that the algorithm gives a success
probability of order 1 in time of order $\sqrt N$, and in $d=4$, we
find a success probability of order $1/\log N$ in time of order
$\sqrt{N \log N}$.  Finally, we investigate the critical point in $d <
4$ and show that the algorithm does not provide substantial speedup.

%%%%%%%%%%%%%%%%%%%%%%%%%%%%%%%%%%%%%%%%%%%%%%%%%%%%%%%%%%%%%%%%%%%%%%%%%%%%%%%
\subsection{Preliminaries}

In this section, we show how the spectrum of $H$ can be understood in
terms of the spectrum of $L$.  An eigenvector of $H$, denoted
$|\psi_a\>$, with eigenvalue $E_a$, satisfies
\be
  H |\psi_a\> = (-\gamma L - |w\>\<w|) |\psi_a\> = E_a |\psi_a\>
\,,
\ee
i.e.,
\be
  (-\gamma L - E_a)|\psi_a\> = |w\>\<w|\psi_a\>
\,.
\label{eq:eigen}
\ee
The state $|\psi_a\>$ is normalized, so $|\<\psi_a|\psi_a\>|^2 = 1$.
Define
\be
  R_a = |\<w|\psi_a\>|^2
\label{eq:Rdef}
\ee
and choose the phase of $|\psi_a\>$ so that
\be
  \<w|\psi_a\>=\sqrt{R_a}
\,.
\label{eq:ra}
\ee
We wish to calculate the amplitude for success,
\be
  \<w|e^{-iHt}|s\> = \sum_a \<w|\psi_a\>\<\psi_a|s\> e^{-i E_a t}
\,,
\label{eq:amp}
\ee
so we only need those $|\psi_a\>$ with $R_a > 0$.

$L$ is the Laplacian of a lattice in $d$ dimensions, periodic in each
direction with period $N^{1/d}$, with a total of $N$ vertices.  Each
vertex of the lattice corresponds to a basis state $|x\>$, where $x$
is a $d$-component vector with components $x_j \in
\{0,1,\ldots,N^{1/d}-1\}$.  The eigenvectors of $-L$ are $|\phi(k)\>$
with
\be
  \<x|\phi(k)\> = {1 \over \sqrt N} e^{i k \cdot x}
\,,
\label{eq:latticevecs}
\ee
where
\ba
  k_j &= {2 \pi m_j \over N^{1/d}} \\
  m_j &= \cond{
    0,\pm 1,\ldots,\pm{1 \over 2}(N^{1/d}-1) & N^{1/d} \mathrm{~odd} \\
    0,\pm 1,\ldots,\pm{1 \over 2}(N^{1/d}-2), +{1 \over 2} N^{1/d} &
    N^{1/d} \mathrm{~even,}}
\ea
and the corresponding eigenvalues are
\be
  \E(k) = 2\left( d - \sum_{j=1}^d \cos\left(k_j\right) \right)
\,.
\ee

Since $\<\phi(k)|w\> \ne 0$, from (\ref{eq:eigen}) we have
\be
  (\gamma \E(k) - E_a) \<\phi(k)|\psi_a\> \ne 0
\ee
for any $k$.  We can therefore rewrite (\ref{eq:eigen}), using
(\ref{eq:ra}), as
\be
  |\psi_a\> = {\sqrt R_a \over -\gamma L - E_a} |w\>
\,.
\label{eq:evec}
\ee
Consistency with (\ref{eq:ra}) then gives the eigenvalue condition
\be
   \<w| {1 \over -\gamma L - E_a} |w\> = 1
\,.
\label{eq:evalgeneral}
\ee
Using (\ref{eq:latticevecs}), this can be expressed as
\be
  F(E_a)=1 \,,\quad
  F(E) = {1 \over N} \sum_{k} {1 \over \gamma \E(k) - E}
\,.
\label{eq:eval}
\ee

\begin{figure}
\includegraphics[width=\columnwidth]{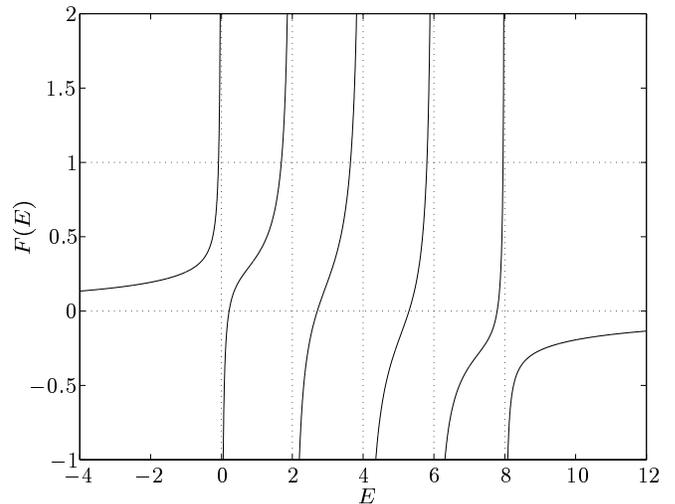}
\caption{The function $F(E)$ for a $d=2$ dimensional periodic lattice
with $N=16$ vertices, at $\gamma=1$.}
\label{fig:f(e)}
\end{figure}

A typical function $F(E)$ is shown in Figure \ref{fig:f(e)}.  This
function has poles where $E=\gamma \E(k)$.  For $E \ne \gamma \E(k)$,
(\ref{eq:eval}) shows that $F'(E)>0$, so there is an eigenvalue of $H$
between every adjacent pair of eigenvalues of $-\gamma L$.  Since
$F(E) \to 0$ as $E \to \pm \infty$, there is also one negative
eigenvalue of $H$ (corresponding to the ground state).  Note that in
the case shown in Figure \ref{fig:f(e)}, the eigenvalues $\E=2,4,6$ of
$-\gamma L$ have degeneracies $4,6,4$ because of the symmetry of the
lattice.  It follows that there are $3,5,3$ eigenvectors of $H$ with
eigenvalues $E_a=2,4,6$, all with $\<w|\psi_a\>=0$ and thus not
relevant to our purpose.  These $11$ eigenvectors, together with the
$5$ relevant ones, make up the necessary total of $16$.

The normalization condition on $|\psi_a\>$ gives
\be
  R_a \<w|{1 \over (-\gamma L - E_a)^2}|w\> = 1
\,,
\ee
i.e.
\be
  R_a = {1 \over F'(E_a)}
\,.
\label{eq:RFprime}
\ee
We also need the overlap of $|\psi_a\>$ with $|s\>$.  Since $L|s\>=0$,
from (\ref{eq:evec}) we have
\be
  \<s|\psi_a\> = -{\sqrt{R_a} \over E_a} \<s|w\>
\,,
\ee
so that
\be
  |\<s|\psi_a\>|^2 = {1 \over N} {1 \over E_a^2 F'(E_a)}
\,.
\label{eq:soverlap}
\ee
Using (\ref{eq:amp}), (\ref{eq:evec}), and (\ref{eq:evalgeneral}),
\be
  \<w|e^{-iHt}|s\>
    = -{1 \over \sqrt N} \sum_a {e^{-i E_a t} \over E_a F'(E_a)}
\label{eq:amplitude}
\,.
\ee
At $t=0$, this gives the sum rule
\be
  \sum_a {1 \over E_a F'(E_a)} = -1
\,.
\label{eq:sumrule}
\ee

We will see that the spectrum of $H$ depends significantly on the
behavior of the sums
\be
  S_{j,d} = {1 \over N} \sum_{k \ne 0} {1 \over [\E(k)]^j}
\,.
\label{eq:sums}
\ee
If $d>2j$, then $S_{j,d}$ can be approximated by an integral
as\footnote{The little-$o$ notation $f(N)=o(g(N))$ means $\lim_{N \to
\infty} f(N)/g(N) = 0$.  In contrast, the more familiar big-$O$
notation $f(N)=O(g(N))$ means there exist constants $c,N_0$ such that
for all $N \ge N_0$, $|f(N)| \le c \, |g(N)|$.}
\be
  S_{j,d} = I_{j,d} + o(1)
\ee
where
\be
  I_{j,d} = {1 \over (2\pi)^d} \int_{-\pi}^{\pi} {\d^d k \over [\E(k)]^j} 
\,.
\ee
The condition $d>2j$ is necessary for $I_{j,d}$ to converge at $k=0$.
The numerical values of $I_{1,d}$ and $I_{2,d}$ for $d \le 10$ are
given in Table \ref{tab:integrals}.  Note that $I_{j,d}$ can also be
calculated using the formula \cite{Mon56}
\be
  I_{j,d} = {1 \over (2d)^j} \int_0^\infty \d\alpha  \,
            {\alpha^{j-1} e^{-\alpha} \over (j-1)!}
            [{\cal I}_0(\alpha / d)]^d 
\ee
where ${\cal I}_0$ is a modified Bessel function of the first kind.

\begin{table}
\begin{tabular}{c@{~~~~}r@{.}l@{~~~~}r@{.}l}
  $d$ & \multicolumn{2}{c@{~~~~}}{$I_{1,d}$} 
      & \multicolumn{2}{c}{$I_{2,d}$} \\ \hline
  3   & 0&253  & \multicolumn{2}{c}{} \\
  4   & 0&155  & \multicolumn{2}{c}{} \\
  5   & 0&116  & 0&0184  \\
  6   & 0&0931 & 0&0105  \\
  7   & 0&0781 & 0&00697 \\
  8   & 0&0674 & 0&00504 \\
  9   & 0&0593 & 0&00383 \\
  10  & 0&0530 & 0&00301
\end{tabular}
\caption{Numerical values of the convergent integrals.  The result for
$I_{1,3}$ is given exactly in \cite{Wat39}; the rest were computed
numerically.}
\label{tab:integrals}
\end{table}

On the other hand, if $d<2j$, then $S_{j,d}$ can be well approximated
by the contribution from values of $k$ small enough that $\E(k)$ is
approximately
\be
  \E(k) \approx k^2 = {(2 \pi m)^2 \over N^{2/d}}
\label{eq:smallk}
\ee
(where we have used the notation $k^2 = k_1^2 + \cdots + k_d^2$).
Then
\be
  S_{j,d} \sim c_{j,d} \, N^{{2j \over d}-1}
\label{eq:highdsum}
\ee
where
\be
  c_{j,d} = {1 \over (2 \pi)^{2j}} \sum_{m \ne 0} {1 \over (m^2)^j}
\,.
\label{eq:csum}
\ee
Here the sum is over all values of the $d$-component vector of
integers $m$ other than $m=0$, and converges for large $m^2$.
Numerically, we find
\be
  c_{2,2} = 0.00664 \,, \quad
  c_{2,3} = 0.0265
\,.
\ee

In the borderline case $d=2j$, $I_{j,d}$ diverges logarithmically at
$k^2$ small and $c_{j,d}$ diverges logarithmically at $m^2$ large.  In
this case
\be
  S_{j,2j} = {1 \over (4 \pi)^j \, j!} \ln N + O(1)
\,.
\label{eq:logsum}
\ee
We will need
\ba
  S_{1,2} &= {1 \over 4 \pi}    \ln N + A + O(N^{-1}) \label{eq:j1d2}\\
  S_{2,4} &= {1 \over 32 \pi^2} \ln N + O(1)
\ea
where $A = 0.0488$ (the case $j=1$, $d=2$ is treated in greater detail
in \cite{Mon69}).

%%%%%%%%%%%%%%%%%%%%%%%%%%%%%%%%%%%%%%%%%%%%%%%%%%%%%%%%%%%%%%%%%%%%%%%%%%%%%%%
\subsection{Phase transition}

In this section, we show that the overlap of the state $|s\>$ on the
ground or first excited state of $H$ exhibits a phase transition at a
critical value of $\gamma$ for any dimension $d$.  In fact, away from
the critical value, $|s\>$ is approximately an eigenstate of $H$, so
Schr\"odinger evolution according to $H$ does not change the state
very much.
In the next section, we will show that the algorithm indeed fails away
from the critical value of $\gamma$, and in the following sections we
will consider what happens near the critical point.

For $\gamma$ larger than the critical value (which will be determined
below), the ground state energy is very close to $0$.  This can
be seen as follows.  The eigenvalue condition (\ref{eq:eval}) for the
ground state energy $E_0$, which is negative, gives
\ba
  1  = F(E_0)
    &= {1 \over N |E_0|} 
     + {1 \over N} \sum_{k \ne 0} {1 \over \gamma \E(k) + |E_0|} \\
    &< {1 \over N |E_0|} 
     + {1 \over N}\sum_{k \ne 0} {1 \over \gamma \E(k)}
    \label{eq:fe0bnd} \\
    &\approx {1 \over N |E_0|} + {I_{1,d} \over \gamma}
\ea
where in the last line we have assumed $d>2$.  In this case, for
$\gamma > I_{1,d}$ (which will turn out to be the critical
value), up to small terms,
\be
  |E_0| < {1 \over N} {\gamma \over \gamma - I_{1,d}}
\,.
\label{eq:groundstate}
\ee
Using (\ref{eq:soverlap}), we have
\ba
  |\<s|\psi_0\>|^2
    &=   \Big[1 + E_0^2 \sum_{k \ne 0}
         (\gamma \E(k)+|E_0|)^{-2}\Big]^{-1} \\
    &> \left[1 + {E_0^2 \over \gamma^2} \sum_{k \ne 0} {1 \over
       [\E(k)]^2}\right]^{-1} \\
    &> 1 - {E_0^2 \over \gamma^2} \sum_{k \ne 0} {1 \over [\E(k)]^2}
\,.
\ea
Inserting the behavior of $S_{2,d}$ from (\ref{eq:sums}),
(\ref{eq:highdsum}), and (\ref{eq:logsum}) and using the bound
(\ref{eq:groundstate}), we find
\be
  1-|\<s|\psi_0\>|^2 < {1 \over (\gamma - I_{1,d})^2} \times
  \cond{O(N^{-1})        & d > 4 \\
        O(N^{-1} \log N) & d = 4 \\
        O(N^{-2/3})      & d = 3 \,. }
\label{eq:groundbiggamma}
\ee
This shows that if $\gamma = I_{1,d} + \epsilon$ for any $\epsilon>0$,
then $1-|\<s|\psi_0\>|^2$ approaches zero as $N \to
\infty$.

If $d=2$, then $I_{1,2}$ is logarithmically divergent, but using
(\ref{eq:j1d2}) in (\ref{eq:fe0bnd}) we can apply a similar argument
whenever $\gamma > {1 \over 4\pi} \ln N + A$, in which case we have
\be
  |E_0| < {1 \over N} {\gamma \over \gamma-{1 \over 4 \pi} \ln N-A}
\label{eq:d2groundstate}
\ee
and
\be
  1-|\<s|\psi_0\>|^2
    < {1 \over (\gamma - {1 \over 4\pi} \ln N - A)^2} \times O(1)
\,.
\ee
This shows that if $\gamma > ({1 \over 4\pi}+\epsilon) \ln N$, then
$1-|\<s|\psi_0\>|^2 \le 1/(\epsilon \ln N)^2$, which approaches zero
as $N \to \infty$.

Similarly, for $d>2$ and for $\gamma < I_{1,d}$, the first excited
state $|\psi_1\>$, with energy $E_1>0$, is essentially $|s\>$.  Here
we find
\ba
  1  = F(E_1)
    &= -{1 \over N E_1} 
       +{1 \over N} \sum_{k \ne 0} {1 \over \gamma \E(k) - E_1} \\
    &> -{1 \over N E_1} 
       +{1 \over N} \sum_{k \ne 0} {1 \over \gamma \E(k)} \\
    &\approx -{1 \over N E_1} + {I_{1,d} \over \gamma}
\,,
\ea
so that, up to small terms,
\be
  E_1 < {1 \over N} {\gamma \over I_{1,d} - \gamma}
\,.
\ee
Again applying (\ref{eq:soverlap}), we find
\be
  1-|\<s|\psi_1\>|^2 < {1 \over (I_{1,d} - \gamma)^2} \times
  \cond{O(N^{-1})        & d > 4 \\
        O(N^{-1} \log N) & d = 4 \\
        O(N^{-2/3})      & d = 3 \,. }
\label{eq:groundsmallgamma}
\ee
We see that $\gamma=I_{1,d}$ is the critical point.  In $d=2$ we can
apply similar reasoning to obtain that for $\gamma < {1 \over 4\pi}
\ln N + A$,
\be
  1-|\<s|\psi_1\>|^2
    < {1 \over ({1 \over 4\pi} \ln N - \gamma)^2} \times O(1)
\,.
\ee
In this case $\gamma = {1 \over 4 \pi} \ln N + A$ is the critical
point.

%%%%%%%%%%%%%%%%%%%%%%%%%%%%%%%%%%%%%%%%%%%%%%%%%%%%%%%%%%%%%%%%%%%%%%%%%%%%%%%
\subsection{Failure of the algorithm away from the critical point}
\label{subsec:failure}

In this section we will show that the algorithm fails away from the
critical point, regardless of dimension.  The results
(\ref{eq:groundbiggamma}) and (\ref{eq:groundsmallgamma}) are actually
sufficient to show that away from the critical point in $d>4$, the
algorithm can be no better than classical search, but we will give a
different argument for consistency of presentation.

First we consider the regime where $\gamma$ is larger than the
critical value.  In the previous section, we saw that in this case,
the ground state energy $E_0$ is small.  This is sufficient to imply
that the success probability is small at all times.  Combining
(\ref{eq:amplitude}) and (\ref{eq:sumrule}), we see that the amplitude
at an arbitrary time must satisfy
\ba
  |\<w|e^{-iHt}|s\>| 
    &\le {1 \over \sqrt N} \left({2 \over |E_0| F'(E_0)} - 1\right) \\
    &\le {2 \over \sqrt N |E_0| F'(E_0)}
\,.
\label{eq:ampbound}
\ea
Furthermore it is clear from the definition of $F(E)$ that
\be
  F'(E_0) \ge {1 \over N E_0^2}
\,,
\ee
so
\be
  |\<w|e^{-iHt}|s\>| \le 2 \sqrt N |E_0|
\,.
\label{eq:bound}
\ee
Using (\ref{eq:groundstate}), we find that for $d>2$,
\be
  |\<w|e^{-iHt}|s\>| \le {2 \over \sqrt N} {\gamma \over \gamma-I_{1,d}}
\,.
\ee
This shows that if $\gamma = I_{1,d}+\epsilon$ for any $\epsilon>0$,
the success probability is never more than a constant factor larger
than its initial value, no matter how long we run the algorithm.  If
$d=2$, then $I_{1,2}$ is logarithmically divergent, but using
(\ref{eq:d2groundstate}) we find
\be
  |\<w|e^{-iHt}|s\>| 
  \le {2 \over \sqrt N} {\gamma \over \gamma-{1 \over 4 \pi} \ln N-A}
\,.
\ee
This shows that the algorithm fails if $\gamma > ({1 \over
4\pi}+\epsilon) \ln N$ for any $\epsilon>0$.

Now we consider the case where $\gamma$ is smaller than the critical
value.  For $d>4$ and $E<0$, we have
\ba
  F(E) &\approx
      {1 \over (2\pi)^d} \int {\d^d k \over \gamma \E(k) + |E|} \\
  &= {1 \over (2\pi)^d} \int {\d^d k \over \gamma \E(k)}
     -{|E| \over (2\pi)^d} \int 
     {\d^d k \over \gamma \E(k) [\gamma \E(k) + |E|]} \\
  &> {I_{1,d} \over \gamma} 
     - {|E| \over \gamma^2 (2\pi)^d} \int {\d^d k \over [\E(k)]^2} \\
  &= {I_{1,d} \over \gamma} - {I_{2,d} \over \gamma^2} |E|
\,.
\ea
Using the fact that $F(E_0)=1$, this shows that
\be
  |E_0| > {\gamma(I_{1,d} - \gamma) \over I_{2,d}}
\,.
\ee
From (\ref{eq:Rdef}) and (\ref{eq:RFprime}), it is clear that $F'(E)
> 1$, so using (\ref{eq:ampbound}) gives 
\be
  |\<w|e^{-iHt}|s\>| < {1 \over \sqrt N} 
                       {2 I_{2,d} \over \gamma(I_{1,d}-\gamma)}
\,.
\ee

A similar argument can be used for $d=3,4$.
With $d=4$, we have
\ba
  F(E) &\approx
      {1 \over (2\pi)^4} \int {\d^4 k \over \gamma \E(k) + |E|} \\
  &= {1 \over (2\pi)^4} \int {\d^4 k \over \gamma \E(k)}
    -{|E| \over (2\pi)^4} \int 
     {\d^4 k \over \gamma \E(k) [\gamma \E(k) + |E|]} \\
  &> {I_{1,4} \over \gamma} - {|E| \over 32 \gamma} 
     \int_0^{2 \pi} {k \, \d k \over {4 \gamma \over \pi^2} k^2 + |E|} \\
  &= {I_{1,4} \over \gamma} - {\pi^2 |E| \over 256 \gamma^2}
     \ln\left(1+{16 \gamma \over |E|}\right)
\,,
\label{eq:4df}
\ea
where the third line follows because $\cos k \le 1-2(k/\pi)^2$ for
$|k| \le \pi$, which implies $\E(k) \ge {4 \over \pi^2} k^2$.  We have
also used the fact that $k^2 \le d \pi^2$ to place an upper limit on
the integral.  This shows that for any $\epsilon > 0$ (with $\epsilon
\le 1$), there exists a $c>0$ such that
\be
  F(E) > {I_{1,4} \over \gamma} 
         - {c |E|^{1-\epsilon} \over \gamma^{2-\epsilon}}
\,,
\ee
so that
\be
  |E_0| > c' \gamma (I_{1,d} - \gamma)^{1/(1-\epsilon)}
\ee
for some $c'>0$, and therefore
\be
  |\<w|e^{-iHt}|s\>| < {1 \over \sqrt N} 
                       {2 \over c' \gamma (I_{1,4}-\gamma)^{1/(1+\epsilon)}}
\,.
\ee
With $d=3$, we have
\ba
  F(E) &\approx
      {1 \over (2\pi)^3} \int {\d^3 k \over \gamma \E(k) + |E|} \\
  &= {1 \over (2\pi)^3} \int {\d^3 k \over \gamma \E(k)}
    -{|E| \over (2\pi)^3} \int 
     {\d^3 k \over \gamma \E(k) [\gamma \E(k) + |E|]} \\
  &> {I_{1,3} \over \gamma} - {|E| \over 8 \gamma} 
     \int_0^\infty {\d k \over {4 \gamma \over \pi^2} k^2 + |E|} \\
  &= {I_{1,3} \over \gamma} - {\pi^2 \over 32 \gamma^{3/2}} \sqrt{|E|}
\label{eq:3df}
\ea
where in the third line we have again used $\E(k) \ge {4 \over \pi^2}
k^2$.
In this case we find
\be
  |E_0| > {1024 \over \pi^4} \gamma (I_{1,3} - \gamma)^2
\ee
which shows that
\be
  |\<s|e^{-iHt}|w\>| < {1 \over \sqrt N}
                       {2 \pi^4 \over 1024 \gamma (I_{1,3}-\gamma)^2}
\,.
\ee

Finally, with $d=2$ we use a different argument.  Here we have
\ba
  F'(E) &\approx
        {1 \over (2\pi)^2} \int {\d^2 k \over [\gamma \E(k) + |E|]^2} \\
  &> {1 \over 2\pi} \int_0^\pi
     {k \, \d k \over (\gamma k^2 + |E|)^2} \\
  &= {\pi \over 4 |E| (|E| + \pi^2 \gamma)}
\label{eq:2dfprime}
\ea
where the second line follows since $\cos k \ge 1-{1 \over 2}k^2$,
which implies $\E(k) \le k^2$.  In the second line we have also used
the fact that the entire disk $|k|\le\pi$ is included in the
region of integration.  Equation (\ref{eq:2dfprime}) shows that
\be
  |E| F'(E) > {\pi \over 4(|E|+\pi^2\gamma)}
\,,
\ee
so that
\be
  |\<w|e^{-iHt}|s\>| < {1 \over \sqrt N}{8(|E_0|+\pi^2\gamma) \over \pi}
\,,
\label{eq:2dbound}
\ee
which is $O(1/\sqrt N)$ for $\gamma=O(1)$, and $O((\log N)/\sqrt N)$
for any $\gamma < {1 \over 4 \pi} \ln N + A$.

The arguments for the case where $\gamma$ is smaller than the critical
value can be made tighter by a more refined analysis.  For example, by
considering the behavior of $F'(E)$, one can give a bound whose
dependence on $I_{1,d}-\gamma$ is linear for all $d>2$, not just for
$d>4$.  Futhermore, the careful reader will note that our bounds for
$d>2$ all become useless as $\gamma \to 0$, but it is easy to see
that the algorithm cannot be successful for small values of $\gamma$.

Altogether, we see that the algorithm cannot work any better than
classical search if $\gamma$ is not chosen close to its critical
value.  It remains to investigate what happens near the critical
point.

%%%%%%%%%%%%%%%%%%%%%%%%%%%%%%%%%%%%%%%%%%%%%%%%%%%%%%%%%%%%%%%%%%%%%%%%%%%%%%%
\subsection{The critical point in \boldmath{$d \ge 4$}}

In this section we investigate the region of the critical point in the
cases where the algorithm provides speedup.  First we consider the case
$d>4$.  Separating out the $k=0$ term in (\ref{eq:eval}), we have
\be
  F(E) = - {1 \over NE} 
         + {1 \over N} \sum_{k \ne 0} {1 \over \gamma \E(k) - E}
\,.
\ee
If $|E| \ll \gamma \E(k)$ for all $k \ne 0$, then for large $N$, we
can Taylor expand the second term to obtain
\be
  F(E) \approx - {1 \over NE} + {1 \over \gamma} I_{1,d}
               + {E \over \gamma^2} I_{2,d}
\label{eq:evalcond}
\ee
which gives
\be
  F'(E) \approx {1 \over NE^2} + {I_{2,d} \over \gamma^2}
\,.
\label{eq:fprime}
\ee
The critical point corresponds to the condition $\gamma = I_{1,d}$.
At this point, setting (\ref{eq:evalcond}) equal to $1$ gives two
eigenvalues,
\be
  E_0 \approx -{I_{1,d} \over \sqrt{I_{2,d} N}} \,, \quad
  E_1 \approx +{I_{1,d} \over \sqrt{I_{2,d} N}}
\,,
\label{eq:criticalenergies}
\ee
which correspond to the ground and first excited state, with a gap of
order $N^{-1/2}$.  Since $\E(k) \approx (2\pi)^2 N^{-2/d}$ for
$m^2=1$,  we see that the assumption $E_0,E_1 \ll \gamma \E(k)$ holds
for all $k \ne 0$.  Furthermore, for the ground and first excited
states at $\gamma = I_{1,d}$, (\ref{eq:fprime}) gives
\be
  F'(E_0) \approx F'(E_1) \approx {2 I_{2,d} \over I_{1,d}^2} 
\,.
\label{eq:criticalfprime}
\ee

Now we want to use (\ref{eq:amplitude}) to compute the time evolution
of the algorithm.  The contribution from all states above the first
excited state is small, since as can be seen using (\ref{eq:sumrule})
we have
\ba
  &-{1 \over \sqrt N} \sum_{E_a>E_1} {1 \over E_a F'(E_a)} \nn
  &\qquad  = {1 \over \sqrt N} \left( 1 + {1 \over E_0 F'(E_0)} 
     +{1 \over E_1 F'(E_1)} \right) \label{eq:sumrule2}
\,.
\ea
Using (\ref{eq:criticalenergies}) and (\ref{eq:criticalfprime}), we
see that the $O(\sqrt N)$ contributions from $1/E_0 F'(E_0)$ and
$1/E_1 F'(E_1)$ cancel, so the right hand side of (\ref{eq:sumrule2})
is $o(1)$.  Thus, using (\ref{eq:amplitude}), we find
\be
  |\<w|e^{-iHt}|s\>| \approx
    {I_{1,d} \over \sqrt{I_{2,d}}} 
    \left| \sin \left({I_{1,d} \, t \over \sqrt{I_{2,d} N}}\right) \right|
\,.
\ee
The success probability is of order $1$ at $t=\sqrt{I_{2,d} N} /
I_{1,d}$.  Straightforward analysis shows that a similar condition
holds so long as $\gamma = I_{1,d} \pm O(N^{-1/2})$, exactly the width
of the region that cannot be excluded based on the arguments of
Section \ref{subsec:failure}.

In $d=4$, $I_{2,d}$ does not converge, so the result is modified
slightly.  In this case (\ref{eq:evalcond}) holds with $I_{2,d}$
replaced by ${1 \over 32 \pi^2} \ln N$, so the ground and first
excited state energies are given by
\be
   E_0 \approx -{I_{1,4} \over \sqrt{{1 \over 32 \pi^2} N \ln N}} \,, \quad
   E_1 \approx +{I_{1,4} \over \sqrt{{1 \over 32 \pi^2} N \ln N}}
\,,
\ee
and we find
\be
  F'(E_0) \approx F'(E_1) \approx {\ln N \over 16 \pi^2 I_{1,4}^2}
\,.
\ee
Therefore
\be
  |\<w|e^{-iHt}|s\>| \approx
      {I_{1,4} \over \sqrt{{1 \over 32 \pi^2} \ln N}}
     \left| \sin 
     \left({I_{1,4} \, t \over \sqrt{{1 \over 32 \pi^2} N \ln N}}\right) 
     \right|
\,,
\ee
which shows that running for a time of order $\sqrt{N \log N}$ gives a
success probability of order $1/\log N$.  Using $O(\log N)$
repetitions to boost the success probability close to $1$, we find a
total run time $O(\sqrt N \, \log^{3/2} N)$.\footnote{In fact, we
could improve the run time of the algorithm to $O(\sqrt N \, \log N)$
using amplitude amplification \cite{BHMT00}.}  One can show that
similar conditions hold so long as $\gamma = I_{1,4} \pm O(\sqrt{(\log
N)/ N})$.

For $d<4$, the expansion (\ref{eq:evalcond}) fails to find states
whose energies satisfy $E \ll \gamma \E(k)$.  Indeed, we will see in
the next section that the algorithm provides no substantial speedup in
these cases.

%%%%%%%%%%%%%%%%%%%%%%%%%%%%%%%%%%%%%%%%%%%%%%%%%%%%%%%%%%%%%%%%%%%%%%%%%%%%%%%
\subsection{The critical point in \boldmath{$d<4$}}

To handle the case $d<4$, we rearrange the eigenvalue condition to
extract the $O(1)$ contribution to $F(E)$:
\be
  F(E) \!=\!
      - {1 \over N E}
      + {1 \over N} \sum_{k \ne 0} {1 \over \gamma \E(k)}
      + {1 \over N} \sum_{k \ne 0} {E \over \gamma \E(k) [\gamma \E(k)-E]}
.
\label{eq:rearranged}
\ee

In $d=3$, we can replace the middle term by $I_{1,3}/\gamma$ for large
$N$.  To explore the neighborhood of the critical point in $d=3$, we
introduce rescaled variables $a,x$ via
\ba
  \gamma &= I_{1,3} + {a \over N^{1/3}} \\
  E      &= {4 \pi^2 I_{1,3} \over N^{2/3}} \, x
\,.
\ea
Since the sum in the third term of (\ref{eq:rearranged}) only gets
significant contributions from small energies, we use
(\ref{eq:smallk}) to give the approximation
\be
  \gamma \E(k) \approx {4 \pi^2 I_{1,3} m^2 \over N^{2/3}}
\,,
\ee
and we can analyze the sum using the same techniques we applied to
calculate $S_{j,d}$ in the case $d<2j$.  Then we have, for large $N$, 
\be
  F(E)
   \approx 1 + {G_3(x)-a \over I_{1,3} N^{1/3}}
\ee
where
\be
  G_3(x) = {1 \over 4 \pi^2} \left(
           \sum_{m \ne 0} {x \over m^2 (m^2 - x)} - {1 \over x}
           \right)
\,.
\label{eq:g3}
\ee
Here the sum is over all integer values of $m$, as in (\ref{eq:csum}),
and similarly converges for large $m^2$.  The eigenvalue condition in
terms of $x$ is $G_3(x) = a$, which has one negative solution $x_0$.
Since $G_3(x)$ is independent of $N$, $x_0$ is independent of $N$, and
the ground state energy $E_0$ is proportional to $N^{-2/3}$.

As we saw in Section \ref{subsec:failure}, a very small ground state
energy implies that the success probability is small at all times.
Using (\ref{eq:bound}), we find
\be
  |\<w|e^{-iHt}|s\>| \le {8 \pi^2 I_{1,3} |x_0| \over N^{1/6}}
\,.
\ee
Therefore the success probability is small no matter how long we run
the algorithm.  This fact is sufficient to imply that the algorithm
cannot produce full square root speedup.  Taking the time derivative
of (\ref{eq:amplitude}), we see that
\be
  {\d \over \d t} |\<w|e^{-iHt}|s\>|
  \le \left|{\d \over \d t} \<w|e^{-iHt}|s\>| \right| 
  \le {1 \over \sqrt N}
\,,
\ee
which implies that
\be
  t \ge |\<w|e^{-iHt}|s\>| \sqrt N
\,.
\ee
Thus the time required to find $w$ using classical repetition of the
evolution for time $t$ is of order
\ba
  {t \over |\<w|e^{-iHt}|s\>|^2}
  &\ge {\sqrt N \over |\<w|e^{-iHt}|s\>|} \label{eq:amptimebound} \\
  &\ge {N^{2/3} \over 8 \pi^2 I_{1,3} |x_0|}
\ea
regardless of $t$.  In other words, the algorithm cannot produce full
speedup.

Similar considerations hold in the case $d=2$.  In this case, the
critical point is at $\gamma = {1 \over 4\pi} \ln N + A$, so we choose
\ba
  \gamma &= {1 \over 4\pi} \ln N + A + a \\
  E      &= {2 \pi \ln N \over N} \, x
\,.
\ea
In this case, we find
\be
  F(E) \approx 1 + {{G_2(x) - a} \over {1 \over 4\pi} \ln N}
\,,
\ee
where $G_2(x)$ is defined as in (\ref{eq:g3}), but with $m$ having two
components instead of three.  Again we find a solution $x_0<0$ that is
independent of $N$, and applying (\ref{eq:bound}) gives
\be
  |\<w|e^{-iHt}|s\>| \le {4 \pi |x_0| \ln N \over \sqrt N}
\,.
\ee
(Note that we could have reached a similar conclusion using
(\ref{eq:2dbound}).)  Using (\ref{eq:amptimebound}), we find
\be
  {t \over |\<w|e^{-iHt}|s\>|^2} \ge {N \over 4 \pi |x_0| \log N}
\,,
\ee
so the algorithm also fails near the critical point in $d=2$.

%%%%%%%%%%%%%%%%%%%%%%%%%%%%%%%%%%%%%%%%%%%%%%%%%%%%%%%%%%%%%%%%%%%%%%%%%%%%%%%
\section{Discussion} 
\label{sec:discussion}

In this paper we have presented a general approach to the Grover problem
using a continuous time quantum walk on a graph.  We showed that
quadratic speedup can be achieved if the graph is a lattice of
sufficiently high dimension ($d>4$).  Although we had originally hoped
to find a fast algorithm in $d=2$, we found that our approach does not
offer substantial speedup in this case.

Our algorithm begins in the state $|s\>$, which is delocalized over
the entire graph.  One might demand instead that we start at a
particular vertex of the graph.  However, it is clear that $|s\>$ can
be prepared from a localized state using $O(N^{1/d})$ local operations.
In fact, we could also prepare $|s\>$ by running the quantum walk search
algorithm backward from a known localized state for the same amount of
time it would take to find $|w\>$ starting from $|s\>$.

The quantum walk search algorithm is related to a search algorithm
using quantum computation by adiabatic evolution.  Adiabatic quantum
computation is a way of solving minimization problems by keeping the
quantum computer near the ground state of a time-varying Hamiltonian
\cite{FGGS00}.  In the adiabatic version of the search algorithm, the
quantum computer is prepared in the state $|s\>$ (the ground state of $H$
with $\gamma$ large), and $\gamma$ is slowly lowered from a large value to
$0$.  If $\gamma$ is changed sufficiently slowly, then the adiabatic
theorem ensures that the quantum computer ends up near the final ground
state $|w\>$, thus solving the problem.  The time required to achieve a
success probability of order $1$ is inversely proportional to the square
of the gap between the ground and first excited state energies.  On the
complete graph, the fact that the gap is only small (of order $N^{-1/2}$)
for a narrow range of $\gamma$ (of order $N^{-1/2}$) means that $\gamma$
can be changed in such a way that time $O(\sqrt N)$ is sufficient to solve
the problem \cite{RC02,DMV01}.  Since the gap has similar behavior for the
hypercube and for $d$-dimensional lattices with $d>4$, quadratic speedup
can also be achieved adiabatically in these cases.  In $d=4$ the gap is of
order $1/\sqrt{N \log N}$ for a range of $\gamma$ of order $\sqrt{(\log
N)/N}$, so the run time is again $O(\sqrt{N} \log^{3/2} N)$.  In $d<4$, no
speedup can be achieved adiabatically.

Yet another way to solve the Grover problem uses a sequence of
measurements of $H$.  For any adiabatic algorithm, there is a related
algorithm that uses only a sequence of measurements to remain in the
ground state of a slowly changing Hamiltonian \cite{CDFGGL02}.  The case
of a hypercube was presented in \cite{CDFGGL02}, and our present
results show that this algorithm can also be used when the graph is a
lattice with $d>4$.  However, to realize the measurement dynamically,
the Hamiltonian $H$ must be coupled to a pointer variable, which must
be represented using auxiliary space.

Although the quantum walk algorithm does not perform as well as the
Aaronson-Ambainis algorithm in $d=2,3,4$, it does have certain
advantages.  The quantum walk algorithm uses simple, time-independent
dynamics rather than a recursive procedure.  Furthermore, the quantum
walk algorithm uses only a single basis state for each vertex of the
graph, whereas the algorithm of \cite{AA03} needs substantial auxiliary
space.

The actual complexity of the search problem in $d=2$ remains an open
question.  It would be interesting either to improve on the algorithms
of \cite{AKR,CG04} or to prove a lower bound showing that full speedup
cannot be achieved.

%%%%%%%%%%%%%%%%%%%%%%%%%%%%%%%%%%%%%%%%%%%%%%%%%%%%%%%%%%%%%%%%%%%%%%%%%%%%%%%
\acknowledgments

We thank Scott Aaronson for discussing his results on quantum search
of spatial regions and for encouraging us to pursue a quantum walk
approach.  We also thank Edward Farhi and Sam Gutmann for numerous
helpful discussions.
AMC received support from the Fannie and John Hertz Foundation.  This
work was also supported in part by the Cambridge--MIT Institute, by
the Department of Energy under cooperative research agreement
DE-FC02-94ER40818, and by the National Security Agency and Advanced
Research and Development Activity under Army Research Office contract
DAAD19-01-1-0656.

%%%%%%%%%%%%%%%%%%%%%%%%%%%%%%%%%%%%%%%%%%%%%%%%%%%%%%%%%%%%%%%%%%%%%%%%%%%%%%%
% References

\end{document}